\documentclass[doublecol]{epl2}

\title{A quantum trampoline for ultra-cold atoms}

\author{M. Robert-de-Saint-Vincent\inst{1} \and J.-P. Brantut\inst{1} \and Ch.J. Bord\'e\inst{2}
\and A. Aspect\inst{1} \and T. Bourdel\inst{1} \and P.
Bouyer\inst{1}}

\shortauthor{M. Robert-de-Saint-Vincent \etal}

\institute{
  \inst{1} Laboratoire Charles Fabry de l'Institut
d'Optique, Univ Paris Sud, CNRS, campus polytechnique RD128, 91127
Palaiseau, France \\
  \inst{2} SYRTE, Observatoire de Paris,
CNRS, UPMC, 61 avenue de l'Observatoire, 75014 Paris, France }


\abstract{We have observed the interferometric suspension of a
free-falling Bose-Einstein condensate periodically submitted to
multiple-order diffraction by a vertical 1D standing wave. The
various diffracted matter waves recombine coherently, resulting in
high contrast interference in the number of atoms detected at
constant height. For long suspension times, multiple-wave
interference is revealed through a sharpening of the fringes. We
use this scheme to measure the acceleration of gravity.}

\begin{document}

\maketitle

\section{Introduction}
Atoms in free fall are remarkable test masses for measuring
gravity, with a host of applications from underground survey to
tests of the equivalence principle \cite{Fray04}. Because of the
quantized character of atom-light interaction, the acceleration of
free falling atoms can be precisely measured with lasers, for
instance by comparing the velocity change of atoms by absorption
or emission of a single photon to the gravity induced velocity
change in a precisely determined time \cite{Clade05}. Furthermore,
atom interferometry exploits the quantum nature of matter-waves
\cite{Cronin09, Berman97, Peters01, Snadden98}. In both cases, an
accurate measurement of gravity demands a long time of free fall,
but it is a priori limited by the size of the vacuum chamber in
which the measurement takes place \cite{Dimopoulos08, Vogel06,
Stern09}. It is possible to overcome this limitation by bouncing
many times the atoms on an atomic mirror \cite{Aminoff93},
realizing a trampoline for atoms \cite{Impens06, Hughes09}. This
scheme can be used to fold the trajectories within a standard
light-pulse atom gravimeter \cite{Hughes09}.

Here we show how to operate a quantum trampoline based on a
periodically applied imperfect Bragg mirror, which not only
reflects upwards the falling atoms, but also acts as a beam
splitter that separates and recombines the atomic wave packets.
This results in multiple-wave \cite{Weitz96, Hinderthur99, Aoki01,
Impens09} atom interference, evidenced by an efficient suspension
of the atoms even though successive leaks at each imperfect
reflection would classically lead to a complete loss of the atoms.
This suspension is obtained at a precise tuning of the trampoline
period, whose value yields directly the local value of gravity
$g$. Our scheme can be generalized to other interferometer
geometries, such as in \cite{Rasel95, Gupta02} replacing perfect
Bragg reflections with imperfect ones.

A classical trampoline for atoms can be experimentally realized by
periodically bouncing them with perfect Bragg mirrors. These
mirrors are based on atom diffraction by a periodic optical
potential \cite{Kozuma99}, i.e. a vertical standing wave of period
$\lambda/2$, where $\lambda$ is the laser wavelength. The
interaction between the atoms and the optical potential leads to
vertical velocity changes quantized in units of $2V_\textrm{R}$,
where $V_\textrm{R}=h/\lambda m$ is the one photon recoil velocity
of an atom of mass $m$ ($h$ is the Planck constant). For long
interaction pulses, the applied potential can be considered as
time independent, and the atom kinetic energy has to be conserved.
This requirement is fulfilled by changing the vertical velocity
component from $-V_\textrm{R}$ to $+V_\textrm{R}$, and vice-versa.
We call this process a resonant velocity transfer. Perfect Bragg
reflection, yielding only resonant velocity transfer, is obtained
by choosing appropriate duration and intensity of the pulse
\cite{Hughes09}. When a perfect Bragg reflection is applied on
atoms freely falling with a velocity $-V_\textrm{R}$, they bounce
upward with a velocity $+V_\textrm{R}$. After a time $T_0 = 2
V_\textrm{R}/g$ ($\approx 1.2$\,ms for $^{87}$Rb), the reflected
atoms have again a velocity $-V_\textrm{R}$ because of the
downwards acceleration of gravity $g$. Repeating this sequence
with a period $T_0$ allows to suspend the atoms at an almost
constant altitude (thick line in Fig.~1). This is a classical
trampoline \cite{Impens06, Hughes09}.

\begin{figure}
\includegraphics[width=0.48\textwidth]{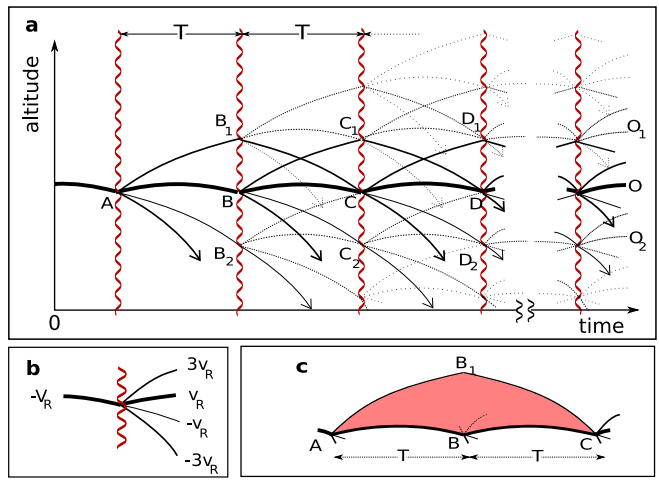} \caption{Atom trajectories in the quantum
trampoline. {\bf a}, atoms, diffracted by periodically applied
imperfect Bragg mirror, explore various paths which eventually
recombine. The probability of a trajectory is represented by the
line thickness. The thick line corresponds to a classical
trampoline associated with perfect Bragg reflections. The arrows
mark the loss channels at $-3 V_\textrm{R}$. {\bf b},  an
imperfect Bragg reflection : an incoming matter-wave with vertical
velocity $-V_\textrm{R}$ is mainly Bragg reflected to
$+V_\textrm{R}$ (thick line). A small fraction is also diffracted
to higher velocities ($+3V_\textrm{R}$ and $-3 V_\textrm{R}$), and
a smaller one transmitted without deviation. {\bf c}, elementary
interferometer\,: from a zero-order trajectory (thick line), an
atom can be diffracted to $+3 V_\textrm{R}$, at point A in the
example shown. It is then Bragg reflected from $+V_\textrm{R}$ to
$-V_\textrm{R}$ at point B$_1$ one period later, and finally
recombines at point C with the zero-order trajectory, thanks to
diffraction from $-3 V_\textrm{R}$ to $+V_\textrm{R}$.}
\label{fig.1}
\end{figure}

To operate the trampoline in the quantum regime, we use imperfect
Bragg reflections, associated with short laser pulses, for which
the kinetic energy conservation requirement is relaxed: Heisenberg
time-energy relation permits energy to change by about $h/\tau$
for a pulse of duration $\tau$. Choosing $\tau \approx
h/4mV_\textrm{R}^2$ allows us to obtain additional velocity
changes from $-V_\textrm{R}$ to secondary diffracted components
with velocities $\pm 3V_\textrm{R}$ (Fig.~1b), hereafter referred
as non-resonant velocity transfers. The matter-wave packet is thus
split into various components that eventually recombine, resulting
in a richer situation where atomic interference plays a dramatic
role (Fig.~1a). For our experimental conditions ($\tau \approx
35$\,$\mu$s), transition from $-V_\textrm{R}$ to $+V_\textrm{R}$
occurs with a probability of 0.93, while the amplitudes $\epsilon$
of the components diffracted to $\pm 3 V_\textrm{R}$ correspond to
a probability  $|\epsilon|^2 \approx 0.03$ ($|\epsilon| \approx
0.17$). The amplitudes of higher velocity components are
negligible, and the probability to remain at $-V_\textrm{R}$ is
0.01. A similar situation occurs for an atom with initial velocity
$+V_\textrm{R}$ : transition to $-V_\textrm{R}$ happens with
probability 0.93 and to $\pm 3 V_\textrm{R}$ with probability
$|\epsilon|^2 \approx 0.03$.

We operate our quantum trampoline as follows. An all-optically
produced ultra-cold sample of $1.5 \times 10^5$ $^{87}$Rb atoms in
the $F=1$ hyperfine level \cite{Clement09} is released from the
trap with a rms vertical velocity spread of 0.1 $V_\textrm{R}$ for
the Bose-Einstein condensate and 0.6 $V_\textrm{R}$ for the
thermal cloud. In this work, the condensate fraction is limited to
0.2. After 600\,$\mu$s ($\approx T_0/2$) of free fall, such that
the mean atomic velocity reaches $-V_\textrm{R}$ because of
gravity, we start to apply imperfect Bragg reflections with a
period $T$, close to the classical suspension period $T_0$. More
precisely, a retroreflected circularly-polarized beam of intensity
4\,mW, 6.3\,GHz red-detuned with respect to the nearest available
atomic transition from $F=1$ is then periodically applied for a
pulse duration $\tau \approx 35$\,$\mu$s. The successive
diffraction events result in several atomic trajectories that
coherently recombine in each output channel, as presented in
Fig.~1a.

\begin{figure}
\includegraphics[width=0.48\textwidth]{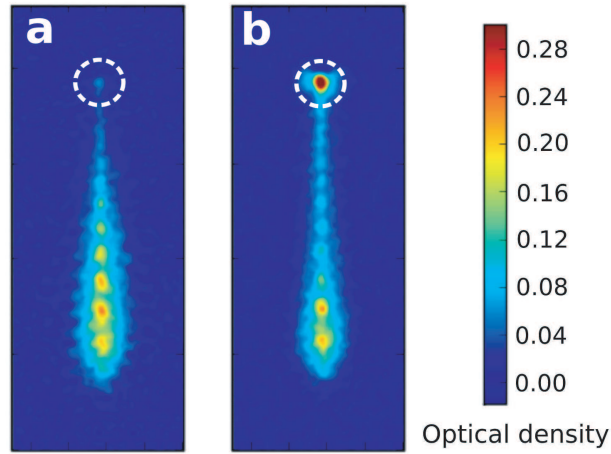} \caption{Outputs
and losses in a 15-pulse quantum trampoline. {\bf a} and {\bf b},
absorption images taken 2\,ms after the last pulse (O in Fig.~1),
in the case of destructive ({\bf a}) and constructive ({\bf b})
interference in the suspended channel. The pulse periods are
$T$=1.206\,ms and $T$=1.198\,ms respectively. The image extends
over 3\,mm vertically. The atoms suspended against gravity lie in
the circle at the top of the images. The spots below correspond to
loss channels and are strongly suppressed in {\bf b} compared to
{\bf a} due to the interference blocking the loss channels. The
two lowest spots correspond to losses at points A and B of
Fig.~1a, which cannot be suppressed by interference.}
\label{fig.2}
\end{figure}

After $N$ pulses,  we wait for a 2\,ms time of flight and detect
the atoms through absorption imaging with resonant light. We
observe distinct wave-packets (Fig.~2). The atoms situated in the
circle at the top have been suspended against gravity, while the
distinct packets below correspond to falling atoms. These atoms
have acquired a velocity $-3 V_\textrm{R}$ after one of the laser
pulses, and then continue to fall, unaffected by the subsequent
pulses. The difference between Fig.~2a and 2b shows that a small
change in the pulse period T results in a dramatic change in the
number of suspended atoms. When suspension is maximum (Fig.~2b),
the losses are strongly suppressed except for the two lowest
spots, which correspond to atoms that have been lost at points A
and B of Fig.~1. This behavior is due to quantum interferences
between the various trajectories as presented in Fig.~1a : for an
adequate pulse period $T$, the interferences are constructive in
the suspended trajectories and destructive in the falling ones,
except at points A and B where no interference can happen. This
blocking of the 'leaking channels' is analogous to the suppression
of light transmission through a multi-layer dielectric mirror.

Our quantum trampoline is a multiple-wave interferometer, where
the fraction of atoms in each output port is equal to the square
modulus of the sum of the amplitudes associated with all
trajectories that coherently recombine at the end. We classify the
contributing trajectories with respect to the number of
non-resonant velocity transfers. The zero-order path is the one
which is reflected from $-V_\textrm{R}$ to $+V_\textrm{R}$ at each
pulse (trajectory ABCD...O in Fig.~1a and Fig.~1c). This is the
path associated with the largest output amplitude (of square
modulus $0.93^N$). The first-order paths are the ones which are
once deviated upwards from the zero-order path, and recombine with
it after twice the period T (for example, the trajectories
AB$_1$CD...O or ABC$_1$D...O in Fig.~1a). All these paths have the
same accumulated interferometric phase that depends on the pulse
period $T$. Their amplitude, proportional to $|\epsilon|^2$, is
small but the number of such paths increases as the number of
pulses $N$. Their total contribution to the probability amplitude
at O scales as $N|\epsilon|^2$. Higher order paths, with more than
2 non-resonant transfers, are less probable individually, but
their number increases faster with the number of pulses. As a
consequence, they can have a major contribution to the final
probability amplitude. More precisely, multiple-wave interference
plays an important role when $N|\epsilon|^2$  becomes of the order
of 1.

\begin{figure}
\includegraphics[width=0.48\textwidth]{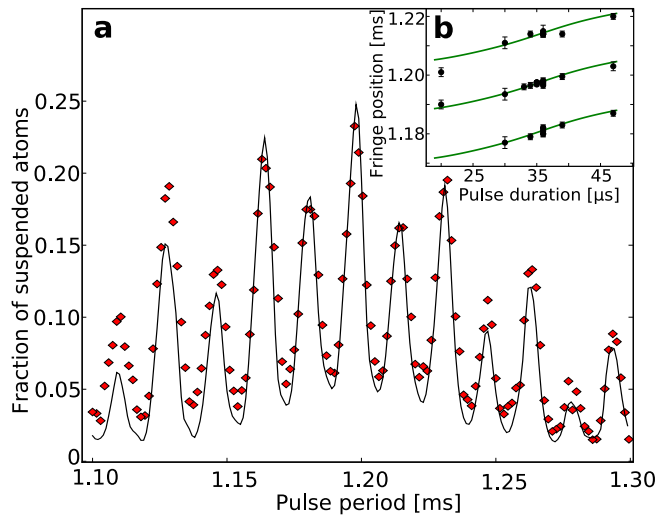} \caption{Interference fringes for a 10-pulse
quantum trampoline. {\bf a}, fraction of suspended atom as a
function of the pulse period. The overall envelope is due to the
velocity selectivity \cite{Impens06} while the modulation is due
to quantum interference. The solid line corresponds to our model
as presented in the text. {\bf b}, position of three consecutive
fringe maxima around the highest maximum, as a function of the
pulse duration, showing the influence of the phase shift $\phi_0$
imprinted by the diffraction pulses : Dots are experimental
points, with error bars reflecting the experimental uncertainties.
Solid lines come from the theoretical model using
g=9.809\,m.s$^{-2}$.} \label{fig.3}
\end{figure}
Figure 3 shows the fraction of suspended atoms for a 10-pulse
quantum trampoline where the interference between the zero- and
first-order paths dominates since $N|\epsilon|^2 \approx 0.3$ is
small compared to 1. When the pulse period T is changed, we
observe interference fringes with characteristic spacing $\Delta T
= 16.6(2) $\,$\mu$s, in agreement with the calculation for the
elementary interferometer (Fig.~1c) \cite{phase}: $\Delta T =
\lambda/4gT$. We also observe an additional modulation with a
fringe spacing of $\Delta T' = 33 $\,$\mu$s, about twice $\Delta
T$. It can be understood by considering the interferometers from A
to O$_1$ and from A to O$_2$, such as AB$_1$C$_1$D$_1$...O$_1$ and
ABC$_1$D$_1$...O$_1$. The corresponding fringe spacing $\Delta T'
= \lambda/4|V_\textrm{R}-gT|$ is equal to $2 \Delta T$ for
$T=T_0$. The output ports O$_1$, O$_2$ of these additional
interferometers are 14\,$\mu$m above or below O. In our absorption
images, we do not distinguish the various ports and the observed
signal is thus the sum of the intensities of the two fringe
patterns. In addition, the total interference pattern is included
in a broad envelop due to the mirror velocity selectivity as
predicted for a classical trampoline in \cite{Impens06} and first
observed in \cite{Hughes09}.

We model our quantum trampoline in a semi-classical approximation
\cite{model}. It makes use of complex amplitudes calculated along
the classical trajectories plotted in Fig.~1a. During each
free-fall, the accumulated phase is given by the action along the
trajectory and, for each diffraction pulse, the matrix of transfer
amplitudes between the various inputs and outputs is calculated by
solving the Schr\"odinger equation in momentum space. At each
output O of the interferometer, we sum the amplitudes of all
possible trajectories from the input A to that output and take the
square modulus to get its probability. For comparison with our
observations, the fraction of suspended atoms is taken as the sum
of the probabilities at all outputs O$_i$. Finally, we take into
account the finite temperature of the initial atomic sample by
summing the results over the distribution of initial velocities.
This model reproduces accurately the whole interference pattern of
Fig.~3a.

\begin{figure}
\includegraphics[width=0.48\textwidth]{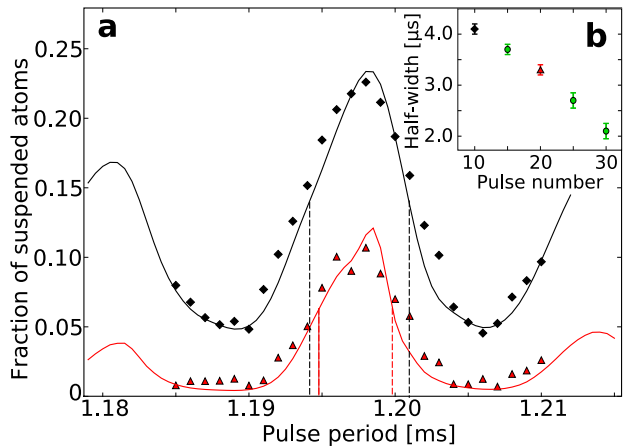}
\caption{Multiple-wave quantum trampoline. {\bf a}, suspended
fraction as a function of the pulse period, for 10 (diamonds) and
20 (triangles) pulses. The contrast evolves from 0.6 to almost 1
and the fringe width is significantly reduced as highlighted by
the vertical dashed lines at half-maximum.  Lines : calculated
suspended fraction, with g = 9.809 m.s$^{-2}$ selected as fitting
parameter. {\bf b},  fringe half-width at half-maximum as a
function of the number of pulses $N$. The narrowing of the fringes
with increasing N is an evidence of the stronger contribution of
higher order paths to the interference pattern.} \label{fig.4}
\end{figure}
When we increase the number of pulses so that $N|\epsilon|^2
\approx 1$, the contribution from higher order paths is not
negligible and we enter a regime of multiple-wave interference.
Fig.~4a shows a comparison of the fringes for the cases of 10 and
20 consecutive pulses. After 20 pulses, we observe a clear
deviation from a sinusoidal pattern, the fringe width decreases
and the contrast increases to almost 1. As plotted in Fig.~4b, the
fringe half-width at half-maximum decreases from 4.1\,$\mu$s after
10 pulses, where $N|\epsilon|^2 \approx 0.3$, to 2.1\,$\mu$s after
30 pulses, where $N|\epsilon|^2 \approx 0.3$. The relative
contributions to the output amplitude at O of zero-, first-, and
second-order paths increase from 1, 0.26, and 0.01 respectively in
the case of 10 pulses to 1, 0.9, 0.32 in the case of 30 pulses.
The finesse of our interferometer, i.e. the ratio of the
full-width at half-maximum of the resonances to the fringe
spacing, is 4 after 30 pulses. This increase of the finesse is an
evidence of the stronger contribution of the higher order paths
when the number N of pulses increases, as expected for a
multiple-wave interferometer \cite{Weitz96, Hinderthur99, Aoki01}.

Our quantum trampoline is sensitive to gravity. From the position
of the broad envelop associated with the classical trampoline, we
can deduce $g$=9.8(1). The same value, with a similar accuracy can
also be inferred from the fringe spacing. However, a measure of
the absolute fringe position allows us to reach a better accuracy.
For this, we need to take into account an additional phase
\cite{phase} $\phi_0$ resulting from the diffraction events, which
varies with the pulse duration (Fig.~3b). We calculate precisely
this phase with our model, and use it to fit the data on Fig.~4a
with $g$ as the only fitting parameter. We find $g = 9.809(4)
$\,m.s$^{-2}$, in agreement with the known value of $g$ in
Palaiseau (9.8095\,m.s$^{-2}$ from WGS84). The uncertainty is due
to our signal-to-noise ratio, and to standing-wave power
fluctuations which affect the complex diffraction amplitudes.
There are several possibilities to improve our setup. First, a
higher number of bounces is achievable, for example starting from
a condensate in a trap with weaker confinement, for which the
velocity spread after release is reduced \cite{Hughes09}. Second,
an adequate shaping of the pulses temporal envelope
\cite{Keller99} could favor well chosen diffracted orders, and
increase the number of contributing trajectories, resulting into a
higher finesse of the fringe pattern. Third, using a standing wave
with a smaller wavelength or atoms with a reduced mass (such as
helium or lithium), the time between bounces would increase and
the precision on $g$ could be improved by several orders of
magnitudes.

\section{Conclusion}
We have presented a quantum trampoline and used it as a
proof-of-principle simple and compact gravimeter, where atoms are
held in a volume of few cubic micrometers. Further investigations
are needed to study the systematic effects and limitations of our
interferometric scheme and to compare it with other compact
sensors \cite{Hughes09, Clade05, Ferrari06}. Our scheme, where the
atomic wave-function is repeatedly split and recombined, is likely
to be weakly sensitive to atom interaction or to laser phase-noise
thanks to averaging over many diffraction events. Beyond the
prospect of miniaturized gravito-inertial sensors, our setup has
potential applications for measuring fundamental forces at small
distances \cite{Carusotto05, Harber05}. It also opens perspectives
for new types of interferometers and new sensor geometries.
Suspended atoms could be used for atomic clock applications
\cite{Impens09} or to build additional interferometers in the
horizontal plane. The interrogation time would then not be limited
by the size of the experimental chamber. The realization of a
multidimensional interferometer measuring simultaneously the
acceleration in three dimensions seems possible \cite{Borde04}.
Our quantum trampoline differs dramatically from its classical
analogue, where the random velocity transfers would result in atom
losses. It provides another clear demonstration of the dichotomy
between classical and quantum dynamics \cite{Moore94, Ryu06,
Kempe09}.

\acknowledgments We acknowledge F. Moron and A. Villing for
technical assistance, R. A. Nyman, J.-F. Cl\'ement and B. Allard
for their work on the apparatus, F. Impens for helpful
discussions. This research was supported by CNRS, CNES as part of
the ICE project, Direction G\'en\'eral de l'Armement, the project
"blanc" M\'elaBoF\'erIA from ANR, IFRAF ; by the STREP program
FINAQS of the European Union and by the MAP program SAI of the
European Space Agency (ESA).


\begin{thebibliography}{0}

\bibitem{Fray04}
  \Name{Fray S., Diez C.A., H\"ansch T.W. \and Weitz M.}
  \REVIEW{Phys. Rev. Lett.}{93}{240404}{2004}.

\bibitem{Clade05}
  \Name{Clad\'e P. \etal}
   \REVIEW{Europhys. Lett.}{71}{730}{2005}.

\bibitem{Cronin09}
  \Name{Cronin A.D., Schmiedmayer J. \and Pritchard, D.E.}
   \REVIEW{Rev. Mod. Phys.}{81}{1051}{2009}.

\bibitem{Berman97}
  \Editor{Berman P.R.}
  \Book{Atom Interferometry}
    \Publ{Academic Press, City}
  \Year{1997}.

\bibitem{Peters01}
  \Name{Peters A., Chung K.Y. \and Chu S.}
   \REVIEW{Metrologia}{38}{25}{2001}.

\bibitem{Snadden98}
  \Name{Snadden M.J., McGuirk J.M., Bouyer P., Haritos K.G. \and Kasevich M.A.}
   \REVIEW{Phys. Rev. Lett.}{81}{971}{1998}.

\bibitem{Dimopoulos08}
  \Name{Dimopoulos S., Graham P.W., Hogan J.M. \and Kasevich M.A. }
   \REVIEW{Phys. Rev. D}{78}{042003}{2008}.

\bibitem{Vogel06}
  \Name{Vogel A. \etal}
   \REVIEW{Appl. Phys. B}{84}{663}{2006}.

\bibitem{Stern09}
  \Name{Stern G. \etal}
   \REVIEW{Euro. Phys. Jour. D}{53}{353}{2009}.

\bibitem{Aminoff93}
  \Name{Aminoff C.G. \etal}
   \REVIEW{Phys. Rev. Lett.}{71}{3083}{1993}.

\bibitem{Impens06}
  \Name{Impens F., Bouyer P. \and Bord\'e C.J.}
   \REVIEW{Appl. Phys. B}{84}{603}{2006}.

\bibitem{Hughes09}
  \Name{Hughes K.J., Burke J.H.T. \and Sackett C.A.}
   \REVIEW{Phys. Rev. Lett.}{102}{150403}{2009}.

\bibitem{Weitz96}
  \Name{Weitz M., Heupel T. \and H\"ansch T.W.}
   \REVIEW{Phys. Rev. Lett.}{77}{2356}{1996}.

\bibitem{Hinderthur99}
  \Name{Hinderth\"ur H. \etal}
   \REVIEW{Phys. Rev. A}{59}{2216}{1999}.

\bibitem{Aoki01}
  \Name{Aoki T., Shinohara K. \and Morinaga A.}
   \REVIEW{Phys. Rev. A}{63}{063611}{2001}.

\bibitem{Impens09}
  \Name{Impens F. \and Bord\'e, C.J.}
   \REVIEW{Phys. Rev. A}{80}{031602}{2009}.

\bibitem{Rasel95}
  \Name{Rasel E.M., Oberthaler M.K., Batelaan H., Schmiedmayer J. \and Zeilinger A.}
   \REVIEW{Phys. Rev. Lett.}{75}{2633}{1995}.

\bibitem{Gupta02}
  \Name{Gupta S., Dieckmann K., Hadzibabic Z. \and Pritchard D.E.}
   \REVIEW{Phys. Rev. Lett.}{89}{140401}{2002}.

\bibitem{Kozuma99}
  \Name{Kozuma M. \etal}
   \REVIEW{Phys. Rev. Lett.}{82}{871}{1999}.

\bibitem{Clement09}
  \Name{Cl\'ement J.-F. \etal}
   \REVIEW{Phys. Rev. A}{79}{061406(R)}{2009}.

\bibitem{phase}
The calculation of the phase difference for the elementary
interferometer (Fig.~1c) leads to $\Delta \phi=\phi_0-4 \pi g
T^2/\lambda$, where $\phi_0$ is a constant resulting from the sum
of the phase shifts acquired during the diffraction events. The
trajectory AB$_1$C...O experiences two non-resonant transfers
while ABC...O experiences none.

\bibitem{model}
Our semi-classical approximation can be justified by decomposing
the sample into a superposition of Heisenberg limited wave-packets
with a momentum spread $\Delta p$ and position spread $\Delta x$
such that (i) $(\Delta p /m) N T < \Delta x$ , (ii) $\Delta p /m
\ll V_\textrm{R}$, and (iii) $\Delta x < 14$\,$\mu$m ($NT$ is the
total duration of the interferometer). In our case, these
conditions are met for $\Delta x \approx 10$\,$\mu$m. During the
free falls, according to condition (i), the expansion of a
wave-packet can be neglected and the exact ABCD formalism
\cite{Borde01, Borde02} reduces to calculating its classical
trajectory (position and momentum) and the classical action along
it. For the pulses, condition (ii) ensures that the momentum
spread is sufficiently low so that the diffraction amplitudes can
be calculated as for plane waves. Finally, condition (iii) implies
that wave-packets ending at different positions do not overlap.
For the condensate part, the previous description is also valid as
the interactions ensure that the initial spatial coherence is lost
because the chemical potential $\mu \approx 1 $\,kHz is such that
$\mu NT/h \gg 1$.

\bibitem{Keller99}
  \Name{Keller C. \etal}
   \REVIEW{Appl. Phys. B}{69}{303}{1999}.

\bibitem{Borde04}
  \Name{Bord\'e C.J.}
   \REVIEW{Gen. Relativ. Gravit.}{36}{475}{2004}.

\bibitem{Carusotto05}
  \Name{Carusotto I., Pitaevskii L., Stringari S., Modugno G. \and Inguscio M.}
   \REVIEW{Phys. Rev. Lett.}{95}{093202}{2005}.

\bibitem{Harber05}
  \Name{Harber D.M., Obrecht J.M., McGuirk J.M. \and Cornell E.A.}
   \REVIEW{Phys. Rev. A}{72}{033610}{2005}.

\bibitem{Ferrari06}
  \Name{Ferrari G., Poli N., Sorrentino F. \and Tino G.M.}
   \REVIEW{Phys. Rev. Lett.}{97}{060402}{2006}.

\bibitem{Moore94}
  \Name{Moore F.L., Robinson J.C., Bharucha C., Williams P.E. \and Raizen M. G}
   \REVIEW{Phys. Rev. Lett.}{73}{2974}{1994}.

\bibitem{Ryu06}
  \Name{Ryu C. \etal}
   \REVIEW{Phys. Rev. Lett.}{96}{160403}{2006}.

\bibitem{Kempe09}
  \Name{Kempe J.}
   \REVIEW{Contemporary Physics}{50}{339}{2009}.

\bibitem{Kinoshita05}
  \Name{Kinoshita T., Wenger T. \and Weiss D.S.}
   \REVIEW{Phys. Rev. A}{71}{011602}{2005}.

\bibitem{Borde01}
  \Name{Bord\'e C.J.}
   \REVIEW{C. R Acad. Sc. - series IV - Physics}{2}{509}{2001}.

\bibitem{Borde02}
  \Name{Bord\'e C.J.}
   \REVIEW{Metrologia}{39}{435}{2002}.

\end{thebibliography}
\end{document}